\documentclass[submission,copyright,creativecommons]{eptcs}
\providecommand{\event}{ICLP 2023}
\usepackage{iftex}

\ifpdf
\usepackage{underscore}         
\usepackage[T1]{fontenc}        
\else
\usepackage{breakurl}           
\fi
\usepackage{amsmath}
\usepackage{graphicx}
\usepackage{multirow}
\usepackage{verbatim} 
\usepackage{fancyvrb}
\newcommand{\janus}{{\it Janus}}

\title{The \janus{} System:\\ Multi-paradigm Programming in Prolog and Python}
            
\author{Theresa Swift
\institute{Johns Hopkins Applied Physics Lab}
\email{theresasturn@gmail.com}
\and 
Carl Andersen
\email{carl.andersen@gmail.com} }

\def\titlerunning{Multi-paradigm Programming in Prolog and Python}
\def\authorrunning{T. Swift \& C. Andersen}

\begin{document}
\maketitle              
\begin{abstract}
Python and Prolog express different programming paradigms, with
different strengths.  Python is wildly popular because it is
well-structured, easy to use, and mixes well with thousands of
scientific and machine learning programs written in C.  Prolog's
logic-based approach provides powerful reasoning capabilities,
especially when combined with constraint evaluation, probabilistic
reasoning, well-founded negation, and other advances. Both languages
have commonalities as well: both are usually written in C, both are
dynamically typed, and both use data structures based on a small
number of recursive types.

This paper describes the design and implementation of \janus{}, a
system that tightly combines Prolog and Python into a single process. \janus{}
bi-translates data structures and offers performance of many
hundreds of thousands of round-trip inter-language calls per second.
Although \janus{} is still new, it has been used in commercial
applications including natural language processing, visual query
answering and robotic automation.  \janus{} was developed for XSB,
but porting \janus{} code to a second Prolog has been
straightforward, indicating that \janus{} is a tool that other
Prologs may easily adopt.

\end{abstract}

\section{Introduction} \label{sec:intro}
Modern Prolog systems offer a rich array of functionality that
reflects a strong research community.  There are well over a dozen
Prolog systems that are actively developed and/or maintained,
including SWI(\cite{WSTL12}), SICStus~(\cite{CarM12}),
Ciao~(\cite{HBCLMMP12}), YAP~(\cite{SCDR12}),
ECLIPSe~(\cite{SchiS12}), GNU Prolog~(\cite{DiAC12}),
Picat~(\cite{ZhKF15}), Trealla,~\footnote{{\tt
  https://github.com/trealla-prolog/trealla}} Tau,~\footnote{{\tt
  http://tau-prolog.org}}
and XSB~(\cite{SwiW12}).  Most of these Prologs are compliant with
Prolog's ISO standard~(\cite{ISO-Prolog}), but also offer diverse functionality 
far exceeding the standard.\footnote{Picat has a non-ISO
syntax, but it is based on the B-Prolog engine (\cite{Zhou12}).} 
We offer several examples. Picat, SICStus, and ECLIPSe are
excellent for constraint-evaluation; Ciao is particularly strong for
program analysis and a flexible syntax; YAP is especially fast for
SLDNF; SWI offers a fast and stable multi-threaded engine; and XSB has
pioneered a wide array of tabling strategies, many of which are now
offered by other Prologs.

Altogether, the dialects of these modern Prolog systems offer numerous
useful features that are either unavailable or rarely available in
other programming languages.  Despite these features
the author's experience is that Prolog is rarely used in industry --
certainly not as often as it might be.  Perhaps the major factor
behind this is the lack of programmers trained in Prolog, but there
are also issues of usability.  The first of these is embeddability.  Many
Prologs can be called from C or Java, but fewer can be called from
other languages popular in industry such as Python or C\#.  Fewer
still can be called from emerging languages such as Rust or
Julia.\footnote{Scryer is an emerging Prolog implemented in Rust.}
While the above problems are more acute for research-oriented Prologs
like XSB than for more heavily used Prologs like SWI or SICStus, the
paucity of packages can make
development in any Prolog uneconomical.  Languages like Java, Python
and Javascript each have package repositories with well over 100,000
packages, perhaps two orders of magnitude more than the total packages
available from currently maintained Prologs.  As a result, the use of
Prolog in a commercial system may be limited to a few modules -- or
most commonly, avoided altogether.

The \janus{} system combines Prolog and Python into a single system
running in a single process, with development possible from either the
Prolog or the Python command line (or IDE).  Because Prolog and Python
are tightly connected, round trip calls between languages can be made
many hundreds of thousands of times per second, so that there is
little need to implement functionality in one language if it is
present in the other.

Several features of Python make it a productive choice of language 
for Prolog integration. Python is currently one of the most popular of all
programming languages both for commercial and scientific computing,
with large package managers that are easy to use.  Like most Prologs,
the most popular version of Python, CPython is written in C, with a
well-written C-API.\footnote{{\tt
  https://docs.python.org/3/c-api/index.html}}  Furthermore, Python is
dynamically typed, as is Prolog.  Nearly all of Python's data
structures are recursively formed using lists, tuples, sets, and
dictionaries; as discussed in Section~\ref{sec:api} these data
structures map well into Prolog's logical terms.

In addition to
establishing a tight, fast connection between Python and Prolog, the
authors put substantial effort towards making \janus{} 's
configuration as easy as possible. \janus{} was originally developed
for XSB and its first version is available as part of the XSB
distribution.  XSB's installation process generates code enabling
either Python and Prolog to load the other language on demand.
Alternately, a test site on the {\tt PyPI} repository allows \janus{}
and XSB to be installed via a Python {\tt pip}
command.\footnote{Installation via the {\tt PyPi} site currently works
only for Linux.}  Once \janus{} is installed, Prolog is at a Python
programmer's fingertips, and vice versa.

In the less than a year since \janus{} ' release, it has been used in
commercial applications including reasoning over large knowledge
graphs, natural language processing, visual query answering, and
situational awareness of autonomous agents. Furthermore, our
experience in porting \janus{} to {\tt LVM} Prolog developed by Permion,
Inc.\footnote{{\tt http://permion.ai}} has shown that \janus{} is easy
for a developer to port. \footnote{Although all functionality
described in this is available in XSB, its first version used
non-generic {\tt xsbpy} and {\tt px} within function and module names.
These names are in the process of being changed.}

This paper focuses on the design and architecture of
\janus{}.  First, Section~\ref{sec:api} describes the design of the
Prolog and Python APIs of \janus{} .  Section~\ref{sec:impl} 
describes implementation of the heart of \janus{} : its bi-translation
between Python and Prolog data structures. This section also discusses 
implementation of calls in the \janus{} interface as well as \janus{}
performance.  Section~\ref{sec:case} lists applications using
\janus{} and describes in detail the implementation of one of a dozen
or so libraries that use \janus{}.\footnote{These applications are
described in more detail in a companion paper about \janus{}
applications \cite{AndS23}.}


\subsection{Related Work}

Most of the Prologs mentioned above are written largely in C and so
offer bi-directional C/C++ interfaces.  Many also offer bi-directional Java
interfaces using JPL (SWI and YAP), Jasper (SICStus), Interprolog
(\cite{Cale04}) (XSB) and other frameworks (including Ciao, YAP,
ECLIPSe and GNU Prolog).  Further, some Prologs also have Python
interfaces (e.g, SWI), Javascript interfaces (e.g., Tau, Ciao) or Rust interfaces (Scryer).
In addition, the Natlog interpreter (\cite{Tarau21}), written in
Python provides a form Prolog-style evaluation within Python.

Three factors distinguish the \janus{} interfaces and how they are
used.  First, \janus{} tightly combines Prolog and Python into a
single process.  As will be discussed in subsequent sections, the
result is that \janus{} is extremely fast. Second, its automatic
bi-translation (Section~\ref{sec:xlate}) allows large and complex data
structures to be copied from one language to another without requiring
special declarations.  Third, compiling and linking is handled by the
Janus configuration process, allowing transparent, on-demand
compilation for both Prolog and Python modules, and on-demand loading
of Python modules when called by Prolog. These features are fully
supported for Python and Prolog modules even when they call other
languages, such as C.

%
\


\section{The \janus{} Programmer's API} \label{sec:api}

This section describes the bi-directional \janus{} API.

\subsection{Calling Python from Prolog}
%
\janus{} is consulted (loaded) by Prolog just as any other module
(currently, XSB and {\tt LVM} Prolog are supported).  When \janus{} is
loaded, Python is also loaded and initialized within the same process
as Prolog; paths to \janus{} and its subdirectories are added both to
the library directories of Python ({\tt sys.path}) and Prolog; and
core Prolog modules are loaded.  Subsequent calls from Prolog cause
the Python interpreter to search for its modules and packages in the same
manner as if Python were stand-alone.

We introduce further \janus{} functionality through a series of
examples.

%
\paragraph{Calling Python Functions}
  The \janus{} Prolog predicates {\tt pyfunc/3} and {\tt pyfunc/4} are
  used to call Python functions.  We invoke the Python standard
  library {\tt json}, which reads and writes JSON strings, to
  illustrate Python calls via {\tt pyfunc/3}.  The Python function
  {\tt json.loads({\em string})} inputs a JSON-formatted string, and
  returns a Python dictionary.  A simple call of {\tt json.loads()} 
  from Prolog and subsequent write
  has the form: {\small
\begin{Verbatim}[xleftmargin=.3in]
?- pyfunc(json,loads('{"name":"Bob", "langs":["English", "GERMAN"]}'),Dict),
   writeq(Dict).
\end{Verbatim}
}
\noindent
and prints out:
{\small
\begin{Verbatim}[xleftmargin=.3in]
pyDict([''(name,'Bob'),''(languages,['English','GERMAN'])])
\end{Verbatim}
}
\noindent
Here, Prolog invokes the Python call
{\small
\begin{Verbatim}[xleftmargin=.3in]
json.loads('{"name":"Bob", "langs":["English", "GERMAN"]}')
\end{Verbatim}
}%
\noindent
which parses the string and outputs a Python dictionary, which is 
then translated into a Prolog term and bound to the {\tt Dict} variable.\footnote{Any errors encountered in the Python execution are caught and wrapped in a Prolog exception.} 

More specifically, \janus{} translates a Python dictionary (a set of
key-value pairs) to a Prolog term {\tt pyDict(List)}, where {\tt List}
is a list of key-value tuples. \janus{} represents each such tuple as
a Prolog term {\tt ''(Key,Value)}, in which the functor {\tt ''} (the
empty string) is chosen to make the pairs resemble Python
tuples.\footnote{The choice of functor {\tt ''} used for tuples is
configurable, as are most \janus{} functors.}  Because Python-Prolog
translations can be complex, multi-level Prolog terms, \janus{}
provides Prolog tools to traverse these data structures that are
analogous to the Python methods {\tt values} and {\tt keys()}.

A slightly different call shows \janus ' support for Python keyword
arguments.  {\small
\begin{Verbatim}[xleftmargin=.3in]
?- pyfunc(json,loads('{"name":"Bob", "langs":["English","GERMAN"]}'),pyDict(L)),
   append(L,[''(gpa,3.5)],L1),
   pyfunc(json,dumps(pyDict(L1)),[sort_keys=1],NewString).
\end{Verbatim}
}
Here, the call to {\tt json.dumps()} is made via the Prolog predicate {\tt pyfunc/4}, which takes 
an additional parameter specifying the optional Python keyword parameter {\tt  sort\_keys}. In this case, 
the output is sorted by key, producing

{\small 
	\begin{Verbatim}[xleftmargin=.3in]
{"gpa": 3.5, "languages": ["English", "GERMAN"], "name": "Bob"}
	\end{Verbatim}
}
\noindent

\paragraph{Using Python Objects and Methods} \janus{} also supports the use of Python 
class objects and methods. We demonstrate these capabilities via calls
to the {\tt fasttext} Python package for Natural Language Processing,
specifically its {\tt lid.176.bin} language identification model
(\cite{FBFJM18}).  The model is loaded by calling an initializer
directly from the {\tt fasttext} package:

{\small
\begin{Verbatim}[xleftmargin=.3in]
pyfunc(fasttext,load_model('./lid.176.bin'),Obj).
\end{Verbatim}
}
\noindent
This call unifies {\tt Obj} with a Prolog reference to the (Python)
model object, represented for this example as {\tt
  pyObj(p0x7faca3428510)}. Using the reference,
the object's class methods can be called, or its class attributes
obtained, via {\tt pydot/[3,4]}.  For example, a call to the class method {\tt predict}:

{\small
\begin{Verbatim}[xleftmargin=.3in]
pydot(pyObj(p0x7faca3428510),predict(
       'Janus is a really useful addition to Prolog! But language detection
        requires a longer string than I usually want to type.'),Lang).  
\end{Verbatim}
}
\noindent
returns the detected language and confidence value.

In the preceding examples, once \janus{} was consulted, Python functions 
and methods were immediately available
to Prolog.  In practice, it is sometimes convenient to write libraries 
containing small amounts of Prolog and Python wrapper code.
For instance, the {\tt jns\_rdflib.P} library transforms rdf triples,
which are maintained as objects in the Python {\tt rdflib} library,
into Prolog terms that can be asserted as facts.  In other cases, it
may make sense to provide an access from Prolog that is a bit higher
level than that provided by Python functions or methods, as will be
illustrated in Section~\ref{sec:case}.




\subsection{Calling Prolog from Python} \label{sec:PyPro}
When used from Python, \janus{} is loaded just like any other Python
package.  When \janus{} is loaded into a Python session, the XSB
executable and its core libraries are loaded and initialized, and
XSB's paths configured. For convenience, Python functions for
compiling and consulting user Prolog code are also provided, as in the
following invocation of Prolog's {\tt consult}: {\small
	\begin{Verbatim}[xleftmargin=.3in] 
>>> import janus
>>> consult(prolog_file) 
	\end{Verbatim}
}

When Prolog calls Python, a goal of \janus{} is to make the Python
data structures and calls as natural as possible for a Prolog
programmer.  Similarly, when Python calls Prolog, the goal is to make
Prolog data structures and calls as ``Pythonic'' as possible.  Several
Prolog features make a Pythonic translation challenging.  Python has
no analog of Prolog's logical variables, nor a direct analog to the
named functors of Prolog terms.\footnote{Using Python's {\tt
  namedtuple} library, classes can be created whose instances
syntactically resemble terms with named functors, but both a class
and its instances have to be specifically instantiated to be used.}
Additionally, the non-determinism of Prolog predicates must be
accounted for; and in Prologs like XSB and SWI that support the
well-founded semantics (WFS), each answer must must be associated with
its truth value, which may be {\em true} or {\em undefined}.

\paragraph{Prolog Deterministic Goals} We first describe how Prolog deterministic goals 
(goals returning exactly one answer) can be called via the \janus{}
Python function {\tt jns\_qdet()}.  Specifically, consider a call from
Python to reverse a list: {\small
\begin{Verbatim}[xleftmargin=.3in]
>>> (Answer,TV) = jns_qdet('basics','reverse',[1,2,3,('mytuple'),{'a':{'b':'c'}}]) 
\end{Verbatim}
}%
\noindent
which calls the Prolog goal
{\small
  \begin{Verbatim}[xleftmargin=.3in]
    basics:reverse([1,2,3,('mytuple'),{a:{b:c}}],Return)
  \end{Verbatim}
}
\noindent
The first two parameters of {\tt jns\_qdet()} are the module ({\tt
  basics}) and the name ({\tt reverse}) of the Prolog predicate to be
called .  The remaining parameters are a (possibly empty) sequence of
ground arguments that are passed to the Prolog predicate: in this case
only one argument is passed, the list to be reversed.  The Prolog
predicate that is called must have one additional argument not
included in the Python call: in this case it is the second argument of
{\tt reverse/2} bound to the variable {\tt Return}.\footnote{A
function {\tt jns\_cmd()} handles similarly restricted goals where no
answer binding needs to be returned.}  Here, the call returns to
Python the tuple {\tt(Answer, TV)}, in which {\tt Answer} is the
binding of the Return variable (the reversed list) and {\tt TV} is the
call's returned truth value (here, {\tt 1}). The \janus{} convention
is that a truth value in Python may be {\tt 1} representing {\em
  true}, {\tt 0} representing {\em false}, or 2 representing {\em
  undefined}. (If a goal fails, {\tt None} is returned for the
answer).\footnote{The version of \janus{} using XSB does not use the
Python keywords {\tt True} and {\tt False} since three truth values
are needed and it was thought best for all truth values to have the
same class.}
{\tt jns\_qdet()} is designed to be
extremely fast, and so is suitable for uses like calling Prolog as
part of a neuro-symbolic loss function for machine learning.

\paragraph{Prolog Non-deterministic Goals} Non-deterministic Prolog goals can be called via the \janus{} Python function \verb|jns_comp()|,
which uses the same parameters as {\tt jns\_qdet()} but also offers
optional keyword parameters.
We introduce a small Prolog knowledge base, {\tt jns\_test} to
illustrate the behavior of {\tt jns\_comp()}:
{\small
\begin{Verbatim}[xleftmargin=.3in] 
test1(a,b,1).                      test1(a,c,2).
test1(a,d,5):- unk(something).     unk(X):- tnot(unk(X)).
\end{Verbatim}
}
\noindent
If {\tt jns\_test} is already loaded in Prolog, then the Python call:

{\small
\begin{Verbatim}[xleftmargin=.3in]
>>>  jns_comp('jns_test','test1','a',vars=2)
\end{Verbatim}
}
\noindent
calls the Prolog goal:

{\small
\begin{Verbatim}[xleftmargin=.3in]
?- jns_test:test1(a,V1,V2).
\end{Verbatim}
}
\noindent
The keyword argument {\tt vars=2} causes a Prolog goal to be formed
with two free variables at the end.  The call returns a list of answer
bindings to {\tt V1} and {\tt V2}, each of which has a distinct truth
value: {\small
\begin{Verbatim}[xleftmargin=.3in]
[((b,1),1), ((c,2),1), ((d,5),2) ]
\end{Verbatim}
}
\noindent
The answer tuple {\tt ((d,5),2)} has an {\em undefined} truth value
because it is proved via the subgoal
\begin{Verbatim}[xleftmargin=.3in]
unk(something)
\end{Verbatim}
    which is {\em undefined} in WFS. Under the hood, XSB's tabled
    evaluation of WFS semantics maintains a {\em delay list} for each
    answer that tracks potential proofs of the answer that are delayed
    until their undefined dependencies are resolved.  In this case,
    the answer {\tt (d,5)} has a non-empty delay list (i.e. its
    undefined dependency is never resolved) \cite{SwiW12}.

{\tt jns\_comp()} has other useful keyword arguments.  The keyword
argument {\tt comprehension=set} returns the answers as a set instead
of a (default) list.  The keyword argument \verb|truth_vals=no_truthvals| drops the truth value from the answer
tuple.  Alternately, {\tt truth\_vals=delay\_lists} causes the full
delay lists of answers to be returned.

\janus{} also supports an alternative means of calling complex Prolog
goals using a Python string to hold the arguments of the goal.  The
Python string is first translated to a Prolog atom, and then the
argument terms are read from the atom.
For example,
a system of numerical constraints could be invoked through a call such
as {\small
\begin{Verbatim}[xleftmargin=.3in]
>>> jns_cmnd('jns_constraints','check_entailed','[[X  > 3*Y + 2,Y>0],[X > Y]]')
\end{Verbatim}
}
\noindent
Here, the predicate {\tt check\_entailed/1} reads the input atom,
sets up its constraints, and calls a constraint evaluator such as {\tt
  clpr}.\footnote{See Volume 2, Section 18.2.4 of the XSB manual.}
In fact, a \janus{} -based interface from
Logtalk (\cite{TheLogtalkHandbook}) expands on this approach by
returning logical variables and their bindings to Python in a
dictionary.

Because {\tt jns\_comp()} returns a list or set, 
it can be used in Python anywhere an iterable object can be
used, for instance in the Python code fragment:

{\small
\begin{Verbatim}[xleftmargin=.3in]
>>> for answer in jns_comp(...):
\end{Verbatim}
} 
\noindent
Although iterable objects are arguably central to Pythonic
programming, the current \janus{} implementation constructs answer
lists/sets eagerly, returning all answers at once, which can cause
performance issues when answer sets are very large. A more incremental
alternative would make use of Python's {\tt generator} type, a
cursor-like mechanism through which answers can be returned lazily via
Python's {\tt yield} statement.  This approach is in fact taken by the
Natlog interpreter, written in Python (\cite{Tarau21}).  However,
multiple generators cannot directly be supported by a single-threaded
Prolog engine since this corresponds to multiple top-level
backtracking points.  Future versions of \janus{} may provide support
for {\em table cursors}, using a lightweight mechanism to backtrack
through completed tables.\footnote{Multi-threaded Prologs can support
multiple backtrack points, but with a likely space overhead, since a
Prolog thread requires more memory than a cursor.}

Finally, \janus{} also supports round-trip callbacks from Python
to Prolog to Python, enabling flexible combinations of the two languages'
capabilities and ecosystems. For example, Prolog reasoning could be leveraged by Python-based
graphical interface libraries like {\tt Tkinter}, web interfaces
libraries like {\tt django}, or IDE plug-ins. Here, round-trip callbacks enable both 
languages to participate in the orchestration of the overall program.


\section{Implementation and Performance} \label{sec:impl}

The \janus{} code base is small: approximately 1500 lines of C code
and 750 lines Prolog code, apart from that used for configuration,
which has made it straightforward to port.  It is also designed to be
as fast as possible.  In this section, we first describe the
implementation of bi-translation, followed by the implementation of
sample \janus{} predicates and functions.
<

\subsection{Bi-Translation of Data Structures} \label{sec:xlate}
Bi-translation is at the heart of \janus{}, which offers one translator
to traverse a Prolog term to create a corresponding object on the
Python heap; and another to traverse a Python object to create a term
on the Prolog heap.

\paragraph{Translating a Prolog Term to a Python Object.}
Prolog to Python translation is performed by the function
\verb|translate_prTerm_pyObj()|
which recursively traverses a Prolog term to convert atoms, integers,
floats, and lists to their Python form. The function also uses
translation conventions described earlier, such as translating terms
having the {\tt ''/n} outer functor to tuples, {\tt pySet/1} to sets,
and {\tt pyDict/1} to dictionaries.  A domain error is thrown if a
term is encountered having some other outer functor.

Figure 1 shows a fragment of C-like pseudo-code for the translation of
{\tt pyDict/1} terms (via the macro {\tt PYDICT\_C}). We note several
points.  First, the fragment begins with the {\tt else if}'
conditional since {\tt translate\_prTerm\_pyObj()} is structured as a
series of conditions rather than as a switch statement.  This choice
was made because the number of possible checks is small, and the
conditions can be ordered by expected frequency. Conditions are
ordered to first check for integers, atoms, floats and then lists. If
the term is a structure, checks are made for tuples, dictionaries and
finally sets.

Second, the code shown largely consists of Python C-API calls that
create and assemble Python objects ({\tt PyDIct\_New()} and {\tt
  PyDict\_SetItem()}) along with calls that interact with the Prolog
heap. In the latter category, ({\tt get\_functor\_string()} returns
the functor of a term as a string, {\tt is\_list()}, {\tt get\_arg()}
returns the $n^{th}$ argument of a structure, while {\tt get\_car()}
and {\tt get\_cdr()} return the head and tail of a list).  When
\janus{} is ported to another Prolog, only the calls that interact
with the Prolog heap need to be redefined. (E.g., by using macros to
redefine functions like {\tt get\_arg()} to corresponding functions in
the target Prolog.)

Third, although {\tt translate\_prTerm\_pyObj()} is recursive,
recursion depth has never been a problem in practice, in part because
iteration is used to traverse through the top-level elements of data
structures as in the while statement of Figure 1.
Furthermore, any Prolog term traversed must correspond to a Python
data structures and in our experience Python data structures like
dictionaries or tuples are rarely nested with depth of more than a
couple dozen.  Instead, lists and sets may have a large number of
elements, and dictionaries a large number of associations at a given
level.

  \begin{figure}[htb] \label{fig:prtopy}
    \hrulefill 
{\small  {\tt
\begin{tabbing}
  foo\=foo\=foo\=foo\=foo\=foo\=\kill
\> else if (get\_functor\_string({\em prTerm}) == PYDICT\_C) \\
\> \>      PyObject {\em pykey}, {\em pyval}; \\
\> \>      prolog\_term {\em list}, {\em elt}; \\
\> \>      {\em list} = get\_arg({\em prTerm}, 1); \\
\> \>      {\em pydict} = PyDict\_New(); \\
\> \>      while (is\_list({\em list}))     \\
\> \> \>	{\em elt} = get\_car({\em list}); \\
\> \> \>	{\em pykey} = translate\_prTerm\_pyObj(get\_arg({\em elt},1)); \\
\> \> \>	{\em pyval} = translate\_prTerm\_pyObj(get\_arg({\em elt},2)); \\
\> \>\>	PyDict\_SetItem({\em pydict},{\em pykey},{\em pyval}); \\
\> \> \>	{\em list} = get\_cdr({\em list}); \\
\> \>      {\em Pyobj} = {\em pydict}; \\
\> \>      return TRUE;
    \end{tabbing}
    }}
  \hrulefill
  \caption{Pseudo-C code for fragment of {\tt translate\_prTerm\_pyObj()}}
\end{figure}

For memory management, Python maintains reference counts of
all objects, and objects with a positive reference count cannot be
garbage collected.  The block of code in Figure \ref{fig:prtopy}
does not need to explicitly decrement Python object reference counts (via {\tt
  Py\_DECREF()}) since object references are automatically
decremented when control exits the end of a code block.\footnote{{\tt
  https://docs.python.org/3/c-api/refcounting.html}}

\paragraph{Translating a Python Object to a Prolog Term.}
Figure 2
shows a fragment of
\begin{Verbatim}[xleftmargin=.3in]
translate_pyObj_prTerm()
\end{Verbatim}
which builds Prolog {\tt pyDict/1} terms from Python dictionaries.
As when translating Prolog to Python, {\tt translate\_pyObj\_prTerm()}
is structured as an ordered series of conditions, where some
conditions call the function recursively.  
The code first sets up the outer {\tt pyDict} functor, and then builds
a list in its first argument, iterating through the dictionary via the
Python C API function {\tt PyDict\_Next()} that sets {\em key} and
{\em value} point to the key and value, respectively, of each
association.  For convenience, a tuple is built with the {\em key} and
{\em value} variables, and {\tt translate\_pyObj\_to\_prTerm()} called
recursively.

One key difference of Python from Prolog is Python's support for
complex classes and class instances.  To translate these to Prolog, a
Python object reference is built and returned to Prolog if the type of
a Python object is not otherwise handled (e.g., for Python binary
objects or complex numbers).

\begin{figure} \label{fig:pytopr}
    \hrulefill 
{\small
    {\tt 
      \begin{tabbing}
foo\=foo\=foo\=foo\=foo\=foo\=\kill
\>  else if(PyDict\_Check({\em pyObj}))  \\
\>\>    prolog\_term {\em head}, {\em tail}; \\ 
\>\>    prolog\_term {\em P} = new\_heap\_cell();\\
\>\>    create\_beap\_functor(PYDICT\_C,1,{\em P});\\
\>\>    tail = get\_arg({\em P}, 1);\\
\>\>    PyObject* {\em tup}; \\
\>\>    PyObject** {\em key}, {\em value}; \\
\>\>    Py\_size\_t* {\em pos} = 0;\\
\>\>    while ( PyDict\_Next({\em pyObj}, {\em pos}, {\em key}, {\em value})) \\
\>\>\>      create\_beap\_list({\em tail});\\
\>\>\>      {\em head} =  get\_car({\em tail});\\
\>\>\>      {\em tup} = PyTuple\_New(2);\\
\>\>\>      PyTuple\_SET\_ITEM({\em tup},0,{\em key});\\
\>\>\>      PyTuple\_SET\_ITEM({\em tup},1,{\em value});		\\
\>\>\>      {\em head} = translate\_pyObj\_prTerm({\em tup});	\\
\>\>\>      {\em tail} = get\_heap\_cdr({\em tail});  \\
\>\>        set\_heap\_nil({\em tail});
\end{tabbing}
}}
    \hrulefill 
    \caption{Pseudo-C code for fragment of {\tt translate\_pyObj\_prTerm()}}
\end{figure}
    
\subsection{Implementation of \janus{} Calls} \label{sec:calls}
This section describes at a high level the implementation of
\janus{} calls using {\tt pyfunc/[3,4]} and {\tt jns\_comp()} as representative examples.

\begin{figure} \label{fig:pyfunc}
{\tt {\small
\begin{tabbing}               
fpoo\=foo\=foo\=fooooooooooooooooooooooooooooooo\=\kill 
1) int pyfunc({\em module},{\em prologTern}) \\
\> PyErr\_Clear() \\
\> {\em modulePtr} = PyImport\_ImportModule{\em (module}) \\
\> if {\em modulePtr} is null throw janus\_error
\> \> \> {\rm /*  module cannot be found */}\\
5) \> {\rm Obtain {\em functor} and {\em arity} from {\em prologTerm}} \\
\> {\em funcPtr} = PyObject\_GetAttrString({\em modulePtr}, {\em  functor})\\
\> if(not PyCallable\_Check({\em funcPtr}))  throw janus\_error \\
\> {\em funcTuple} = PyTuple\_New({\em arity})\\
\> for(int {\em i} = 1; {\em i} $<$= {\em arity}; {\em i}++) \\
10) \>\> Set the translated $i^{th}$ argument of {\em funcTuple} to the
$i^{th}$ argument of {\em prologTerm} \\
\> Set up dictionary of keyword arguments if called by {\tt pyfunc/4}
\\ \> {\em retValue} = PyObject\_CallObject({\em funcPtr}, {\em
  funcTuple},{\em dictionary}) \\ \> if ({\em retValue} == NULL) throw
janus\_error \\ \> return translate\_pyobj\_prTerm({\em retValue})
\end{tabbing}
}}
    \hrulefill 
    \caption{Pseudo-C code for {\tt pyfunc/[3,4]}}
\end{figure}
\paragraph{Calling Python from Prolog: pyfunc/[3,4].}
%
Figure 3 
provides
pseudo-code for {\tt pyfunc/[3,4]} (i.e.,
with and without keyword arguments).  Much of the work is done by {\tt
  translate\_prTerm\_to\_pyObj()} to translate the call, and {\tt
  translate\_pyObj\_to\_prTerm()} to return the results.
The
pseudo-code provides a taste of how Python's C API is used. The Prolog
module, functor name, and (if applicable) the list of keyword
arguments are translated to Python objects via C API calls.  The
module loading in lines 2-3 of Figure \ref{fig:pyfunc} has the benefit
that Python modules are loaded {\em on-demand}: if the goal {\tt
  pyfunc({\it module},...)}  is called, {\em module} will be loaded if
necessary without needing any special import declarations. Note that
since Python functions are variadic, the function name can be resolved
in line 6 without specifying its cardinality.
%
To properly handle errors, the Python error buffer is cleared in line
1; any Python error ``caught'' in line 13 is re-thrown as a Prolog
error containing the Python error message and backtrace in addition to the
Prolog message and backtrace.  

\paragraph{Calling Prolog from Python: jns\_comp.}
%
%
The implementation of a call
{\small
\begin{Verbatim}[xleftmargin=.5in]
jns_comp(module,functor,arg_1,...,arg_n,varnum=m)
\end{Verbatim}
}
\noindent
consists of a C function and a Prolog predicate.  First the C function
creates the goal
{\small
\begin{Verbatim}[xleftmargin=.5in]
jns_comp(PrologGoal,Flag,Comp)
\end{Verbatim}
}
 on the Prolog heap, where {\tt PrologGoal} is the goal 
{\small
\begin{Verbatim}[xleftmargin=.5in]
module:functor(arg_1,...,arg_n,var_1,...var_m)
\end{Verbatim}
}
\noindent
created from the arguments of {\tt jns\_comp()} as described in
Section~\ref{sec:PyPro}. Here, {\tt Flag} is an integer encoding of 
keyword arguments, and {\tt Comp} is unified with the list or set
constructed by Prolog.
Due to space constraints, we present in detail only the Prolog
predicate {\tt jns\_comp/3}. 

\begin{figure}[tb] \label{fig:jnsComp}
{\sf {\small
\begin{tabbing}               
foo\=fooo\=foooooooooooooooo\=ooooooooo\=foo\=foo\=foo\=\kill 
1) jns\_comp({\em PrologGoal},{\em Flag},{\em Result})\>\>\>\>{\rm /* Pseudo-Prolog code */}\\
\> If no\_truthvals({\em Flag}) \\
\> \> findall({\em Binding},({\em PrologGoal},strip\_bindings({\em PrologGoal},{\em Flag},{\em Binding}),{\em InterimResult}) \\
 \> Else if tabled({\em PrologGoal}) \\
5)\> \> findall({\em Binding},({\em PrologGoal},table\_bindings({\em PrologGoal},{\em Flag},
                       {\em Binding}),{\em InterimResult}) \\
\> Else abolish\_table\_pred(jns\_table\_goal(\_)) \\
\>\> findall({\em AnswerTuple},(jns\_table\_goal({\em PrologGoal}),\\
\>\>\> table\_bindings({\em PrologGoal},{\em Flag},{\em AnswerTuple}),{\em InterimResult}) \\
\> If set\_comp({\em Flag}) {\em Result} = pySet(sort({\em InterimResult})) else {\em Result} = {\em InterimResult}
\end{tabbing}
}}
    \hrulefill 
    \caption{Pseudo-Prolog code for {\tt jns\_comp()}}
\end{figure}

Since {\tt jns\_comp()} is treated as a list or set comprehension in
Python, the Prolog answers must be gathered into a list or set.  As
shown in Figure 4,
if no truth values are needed, this can be done directly through {\tt
  findall/3}.  In this case each answer tuple
(Section~\ref{sec:PyPro}) consists solely of answer bindings.
Within the findall, after each success of {\tt PrologGoal} with
substitution $\theta$, the answer bindings must be extracted from the
last {\em varnum} arguments of {\tt PrologGoal}$\theta$.

If truth values are to be returned along with the answer bindings,
tabled calls may need to be used to retrieve undefined results.  If
{\tt PrologGoal} isn't a tabled predicate, the findall calls {\tt
  PrologGoal} through the tabled meta-call {\tt jns\_table\_goal/1},
otherwise {\tt PrologGoal} can be called directly.  In lines 5 and 7
the findall first calls {\tt PrologGoal} and then backtracks through
the completed table using the predicate {\tt get\_residual/2} which
binds {\tt PrologGoal} with each answer substitution $\theta$ and
returns the answer's delay list.  Answer bindings are taken from {\tt
  PrologGoal}$\theta$ as in the non-tabled case.  The answer tuple is
constructed from these bindings along with either a numeric truth
value or the delay list of the answer.  If set comprehension is
required, the list of tuples is sorted and then clothed in {\tt
  pySet/1} which signals to the translation that a set is to be
returned to Python.

\section{Performance} \label{sec:perf}
Initially, \janus{} was coded using the Python {\tt ctypes} package,
which did not provide adequate performance, so Python's lowest level
C-API was used instead.  Table~1 shows the average times for both
intra- and inter-language calls.\footnote{All benchmarks were run
within a VM on a server with an AMD EPYC 7542 32-Core Processor
running Ubuntu 20.04, using Python 3.9.16 and the repo version of XSB
of February, 2023, available at {\tt xsb.sourceforge.net.  }
The smallest time over three runs was taken for each benchmark, but
since these tests were done on a shared server all times should be
regarded as approximate.  Benchmark tests used in this paper are
available from the XSB repo version under {\tt
  xsbtests/python\_tests}.}  As an aside, for standard Prolog
benchmarks the performance of XSB is neither the fastest nor the
slowest of the Prologs listed in Section~\ref{sec:intro}.

Table~1 compares the time within and between each
language.  The first row shows the time per iteration of a simple loop
that calls a function or predicate to decrement a counter.  The second
row calls the haversine function,
\footnote{{\tt
  https://en.wikipedia.org/wiki/Haversine\_formula}}
a small function that makes multiple trigonometric and exponentiation
calls.  The third row tests creation of 20 element lists, via list
comprehension based on a set of 20 elements in Python and {\tt
  findall/3} calling a predicate in XSB.  From these timings, XSB is
generally faster than native Python, except for list
comprehension/findall: this is likely due to the difference between
calling a predicate and reading a list from a set.  Table~1 also shows
that for Python calling XSB, while there is an overhead for
inter-language calls, this overhead is still small: many hundreds of
thousands of round-trip Prolog/Python calls can be made per second.
Even for the small haversine function, Prolog calling Python is only
about 3x slower than a Python-only loop.


\begin{table}[hbt] \label{tab:calls}
\centering
{\small
\begin{tabular}{r|r|r|r|r}  
              & XSB Only & Python Only & XSB/Python & Python/XSB\\
  Simple Loop & 30 ns       & 78 ns      & 1.98 $\mu$s & 79 $\mu$s \\ 
  Haversine   & 675 ns      & 1.5 $\mu$s & 5.07 $\mu$s & 81 $\mu$s  \\ 
  List Comp.  & 92 $\mu$s   & 586 ns     & 5.69 $\mu$s & 584 $\mu$s \\ 
\end{tabular}  
}
\caption{Timings of Intra- and  Inter-languaage Calls}
\end{table}

Python calling XSB is slower than XSB calling Python.  For the first
two rows, timings are reported for {\tt jns\_qdet()} (the timings for
{\tt jns\_cmd()} are nearly the same).  The third row uses {\tt
  jns\_comp()} with its default values, which incurs an overhead due
to need to use {\tt findall/3}, to return only the bindings of calls,
as described in the previous section.  Surprisingly, tests that set
{\tt truth\_vals=None} did not run significantly faster than the
default.  Nonetheless, even the slowest timing in Table \ref{tab:calls} is still
fairly fast, supporting thousands of round trips calls per second.

The timings in Table \ref{tab:calls} transferred small amounts of data between the
languages.  To illustrate the performance of bi-translation, Table \ref{tab:xfer}
shows the transfer time per element of tuples, lists, sets and
dictionaries of varying sizes, each of which consisted only of
integers.  For these tests, XSB called the Python function 
{\small
\begin{Verbatim}[xleftmargin=.5in]
def bitranslate(X):
   return(X)
\end{Verbatim}
}
\noindent
so that Table \ref{tab:xfer} represents the average of times to translate a data
structure from Prolog to Python and from Python to Prolog.  Because
XSB allows a maximum arity of 64K for predicates, the last two columns
for tuples were not tested.  While there was unavoidable variance in
the timings, Translation times for sets and dictionaries, which in
Python are based on hash tables, are sometimes higher than for lists
and tuples.  Nonetheless, the cost of per-element transfer is always
small.

\begin{table}[hbt] \label{tab:xfer}
	\centering
  {\small 
\begin{tabular}{r|r|r|r|r|r|r}  \hline 
      & 10 elts & 100 elts & 1000 elts & 10000 elts & 100000 elts & 1000000 elts\\
Tuple & 30     & 28.9    & 36.1     & 48.7      & *          & *          \\
List  & 53.8   & 54.3    & 73.4     & 58.4      & 77.8       & 73.5       \\
Set   & 109    & 104     & 78.1     & 85.5      & 62.0       & 86.1       \\
Dict. & 113    & 40      & 221      & 58.5      & 90.5       & 72.5       \\  \hline
\end{tabular}
}
\caption{Per-element round-trip transfer cost for data structures of increasing size. All times in ns.}
\end{table}

Python's garbage collection works by maintaining reference counts to
each object: only when an object's reference count is 0 can the object
be safely abolished and its space reclaimed.  Because of \janus{}' use
of bi-translation, it does not create persistent references to Python
objects, except when a Python object reference is returned to Prolog,
for use in e.g., {\tt pydot/4}.  Otherwise, the bi-translation code
  of \janus{} is able to execute without causing a memory leak due to
  stray references.  This has been verified by tests of {\tt
    jns\_cmd()}, {\tt jns\_qdet()} and {\tt jns\_comp()} where the
  heap analysis toolset {\tt guppy.heapy}\footnote{{\tt
    https://pypi.org/project/guppy3}} was used to check for
  uncollected Python objects.


\section{Libraries} \label{sec:case}

To support the use of \janus{} in applications, we have steadily built a library of \janus{} interfaces to
Python packages.  The XSB distribution includes interfaces to Elsasticsearch\footnote{{\tt
  https://www.elastic.co}}, Faiss (Facebook AI Similarity
Search)\footnote{{\tt
  https://engineering.fb.com/2017/03/29/data-infrastructure
}} fastText word vectors (\cite{FBFJM18}), the GIS library
Geopy\footnote{{\tt https://geopy.readthedocs.io/en/stable}}, Google
Maps\footnote{{\tt https://developers.google.com/maps}}, Google
Translate\footnote{{\tt https://cloud.google.com/translate}}, RDFlib
(for {\tt hdt}, {\tt jsonld}, {\tt ntriples}, {\tt nquads}, and {\tt
  turtle})\footnote{{\tt https://rdflib.readthedocs.io/en/stable}},
shapely (an interface to GIS files)\footnote{{\tt
  https://shapely.readthedocs.io/en/stable/manual.html}},
SpaCy\footnote{{\tt https://spacy.io}}, Wikidata-hdt\footnote{{\tt
  https://www.rdfhdt.org/datasets}} , Wikidata
Integrator\footnote{{\tt
  https://github.com/SuLab/WikidataIntegrator}}and others.

Because of the mapping of data structures between Python and Prolog,
some of these interfaces need not be large.  For instance, the {\tt
  jns\_elastic} library contains predicates to create and search
Elasticsearch indices in various ways.  Because the Python interface
to Elasticsearch relies on sending and receiving JSON structures over
a simple REST interface, much Elasticsearch functionality can be
called directly using {\tt pyfunc/[3,4]} or {\tt pydot/4}.  As an
example, given a Python connection {\tt conn}, searching an
Elasticsearch index {\tt ind} via a search condition {\tt bdy}
(represented as a Python dictionary) can be done via the Python
statement {\small
\begin{Verbatim}[xleftmargin=.5in]
>>> result = conn.search(index=ind, body=bdy)
\end{Verbatim}
}
\noindent
For a Prolog connection reference {\tt Conn}, index atom {\tt Ind} and
Prolog dictionary term {\tt Bdy}, the analogous statement is quite
similar:
{\small
\begin{Verbatim}[xleftmargin=.5in]
pydot(Conn.search(),[index=Ind, body=Bdy],Result)
\end{Verbatim}
}
\noindent
Indeed, a few of the \janus{} libraries are little more than starting
examples showing how common calls to a Python package are formed in
\janus{}.

\paragraph{Case Study of a Library: jns\_spacy.} Some \janus{} libraries,
such as {\tt jns\_spacy}, implement useful Prolog functionality on top
of the basic Prolog-Python interface.  SpaCy is a widely used Python tool that
exploits neural language models to analyze text via tools such as
dependency parses, named entity recognition, relation extraction, and
sentence or span vectors based on Roberta~(\cite{LOGDJCLLSZS19}) or
other transformer-models.

To use {\tt jns\_spacy}, a SpaCy model is first loaded via the Prolog convenience predicate:
{\small
\begin{Verbatim}[xleftmargin=.5in]
?- load_model(en_core_web_trf). 
\end{Verbatim}
}
\noindent
  which loads SpaCy's English transformer model.  Multiple models can be
loaded when text from more than one language must be
analyzed. Once a model is loaded, analytics commands can
be called:
{\small
\begin{Verbatim}[xleftmargin=.5in]
?- proc_string(en_core_web_trf,'She was the youngest of the two daughters 
of a most affectionate, indulgent father; and had, in consequence of 
her sister's marriage, been mistress of his house from a very early period.',Doc)
\end{Verbatim}
}
\noindent
The above command uses {\tt en\_core\_web\_trf} to analyze the atom in the second
argument, returning a Python object reference {\tt Doc} to the SpaCy result.
Text files can be analyzed in a similar manner.  At this stage,
either the {\tt Doc} object can be queried directly, or the object's
dependency graph can be asserted into Prolog via
{\small
\begin{Verbatim}[xleftmargin=.5in]
?- token_assert(Doc).
\end{Verbatim}
}
\noindent
The Prolog form of the dependency graph has binary relations
connecting nodes of the form
{\small
\begin{Verbatim}[xleftmargin=.5in]
token_info(Index,Text,Lemma,Pos,Tag,Dep,EntType)
\end{Verbatim}
}
Such a node contains in order: the character offset of the token within
the string, the token itself, the lemma of the token (the base form of
a verb or singular form or a noun), the token's part of speech, the
fine-grained part of speech, the tokens relation with it's parent in
the dependency graph, and the entity type if any, for a span containing
the token.

A highly simplified example gives a taste of how Prolog can exploit 
the above interface to better use the results of neural models.  Here, a Prolog rule 
infers a spatial proximity relation from the span information in the dependency graph:
{\small
\begin{Verbatim}[xleftmargin=.5in]
trigger_relation(class(proximity),Span,rel(SibSpan,ChildSpan),[spatial]):-
    dg_subord_conj(Span,SibSpan,ChildSpan),
    span_EREid(SibSpan,physical),
    span_EREid(ChildSpan,physical).
\end{Verbatim}
}
\noindent
The rule infers spatial proximity between entities (here represented
as entity textual spans) {\tt Span} and {\tt ChildSpan}.  The rule
infers proximity if (i) {\tt Span} has a dependency graph sibling {\tt
  SibSpan} that is a physical entity (e.g., person, place, object);
and (ii) {\tt SibSpan} is the head of a subordinating conjunction to
another entity {\tt ChildSpan}.  Both entities are required to be of
the class {\tt physical} by {\tt span\_EREid/2} which checks the
classes of relation and event ids as well as entity ids.  Other
clauses (not shown) represent the different forms of a subordinating
conjunctions using the Prolog form of the SpaCy dependency graph.

The above rule is a simplified example of actual rules used in an
textual analytic application assessing how changing information
affected the behavior of human agents.  In our work, we use rule-based
reasoning as a complement to neural models for neural extraction and
annotation (e.g. SpaCy).  Neural models are the standard for entity
and relation extraction, but their results are often noisy, especially
if they were trained on data different than the run-time domain.
Accordingly rules like the above are useful to refine the meaning of
neural models, to map a model trained on one ontology to a different
ontology, to disambiguate the results of relation extraction (perhaps
using defeasibility), and to assign the extraction a useful
likelihood.
\footnote{The direct use of certainty
	measures from many neural models is unreliable, since the softmax
	layer of neural models may artificially reduce the entropy of its
	output distribution.}
While we have not done so, similar rules could also be used for weak supervision of neural models using
data labeling, as in the Snorkel framework.\footnote{{\tt https://snorkel.ai/}}

\section{Discussion}

As described in a companion paper (\cite{AndS23}), while \janus{} is
still new, it has formed an essential basis for several commercial and
institutional applications using XSB Prolog; it has also formed the
basis for several commercial applications using {\tt LVM} Prolog.  The
largest of the XSB applications involved reasoning over a large scale
knowledge graph (${\cal O}(10^8)$ edges) of entities, relations and
events created by analyzing thousands of documents, images and videos.
In this applications, XSB both performed reasoning, and orchestrated
the integration of various Python tools, such as SpaCY and FastText,
along with external KBs (e.g. Wikidata) accessed via Python.
Orchestration was also a feature of a commercial research project in
which Ergo, (\cite{GKSFB23} a system based on XSB, orchestrated and
reasoned about natural language analysis and geo-spatial data; as well
as a feature of a separate commercial research project on visual query
answering.  While these applications used Prolog calling Python, a
more recent project used Python bindings to the Robot Operating System
(ROS)\footnote{{\tt https://www.ros.org}} to call XSB to perform
situational assessment reasoning for an autonomous agent.

This small flurry of applications is no coincidence.  Prolog's
reasoning power and scalability has advanced over the last two
decades: Prolog systems now include high-quality finite-domain and
numerical constraint systems, event-action rules, defeasible logics
based on the well-founded semantics with priorities and integrity
constraints, and probabilistic and T-norm based reasoning.  Yet one
may argue that the application of these reasoning methods has been
throttled by a lack of data.  Most Prolog systems provide interfaces
to relational databases, but few interface to the current generation
of no-SQL databases such as Elasticsearch or MongoDB, or to systems
like Faiss or Annoy that store high-dimensional numerical knowledge
vectors. Most Prolog systems can be called by a language like C or
Java, but it is not always easy to leverage these interfaces to embed
Prologs in frameworks like ROS.  Indeed, the types of data that
applications need and the types of frameworks that applications use is
constantly changing.  The ability to read any new form of data or fit
into any new framework is only possible for the most heavily used
languages like Java, C\#, Javascript and Python.  However, the use of
the \janus{} framework has allowed two Prologs to view Python as a
term-server for nearly any kind of data, and for Python to view Prolog
as a reasoning mechanism working over its standard data types.

{\small
{\bf Acknowledgements} The authors would like to thank Michael Kifer
for huge help with \janus{} configuration, David Warren for porting the
Prolog to Python bridge to Windows, Muthukumar Suresh who contributed
code used in the Prolog to Python bridge, and Paulo Moura who answered
numerous questions about different Prologs.
}

%
\bibliographystyle{eptcs}
\bibliography{shortstring,all}
\end{document}